\documentstyle[12pt]{article}
\begin{document}
\font\bss=cmr12 scaled\magstep 0
\title{Rational soliton on the plane wave background for the
(1+2) $|\phi|^4$-model with negative coupling}
\author{ Alla A. Yurova* and Artyom V. Yurov**
\small\\ $*$Mathematical Department \small\\Kaliningrad Technical
University, Russia\small\\ yurov@freemail.ru \small\\
$**$Theoretical Physics Department, \small\\ I. Kant University of
Russia \small\\  artyom\_yurov@mail.ru}
\date {}
\renewcommand{\abstractname}{\small Abstract}
\maketitle
\maketitle
\begin{abstract}
We show that $(1+2)$ nonlinear Klein-Gordon equation with negative
coupling admits an exact solution which appears to be the linear
superposition of the plane wave and the nonsingular rational soliton.
\end{abstract}
\thispagestyle{empty}
\medskip
\section{Introduction}

The nonlinear Klein-Gordon equation ($\phi^4$- or
$|\phi|^4$-models) is one of those equations that frequently arise
in seemingly different physical applications \cite{Bishop-80},
\cite{Makhankov}, \cite{Kudr}, \cite{Aubry}, \cite{Bogol-Makh-77}.
Unfortunately, this equation is nonintegrable. In the simplest
case of $(1+1)$ it is relatively easy to find out the stationary
solutions (kinks for the $\phi^4$-model, as a particular example).
In case of general position $(1+D)$, however, there exist no
general methods of construction of the exact solutions of
Klein-Gordon equation - bar the case, when the solutions by
definition possess enough symmetries. Such symmetries usually
result in possibility of reduction of the problem into the
one-dimensional case, allowing for complete integration of the
initial equation. For example, in the massless theory with
potential
$$
V(\phi)=-\frac{\lambda\phi^4}{4},
$$
the Euclidean solution with the $O(4)$ symmetry can be obtained upon the introduction of
the proper boundary conditions. As the result one will get the so-called Fubini
instanton: one-parameter nonsingular solution $\phi(r)$
($r=\sqrt{\sum^{4}_{i=1} {x_i^2}}$, the $x_i$ are the Euclidean coordinates) with
finite Euclidean action \cite{Fubini}. What is somewhat surprising is the fact
that for any nonvanishing $m$ and the potential of the form
$$
V(\phi)=\frac{m^2\phi^2}{2}-\frac{\lambda\phi^4}{4},
$$
the instanton solutions does not exist. This tiny example gives but a glimpse of what highly nontrivial properties does the solutions of the nonlinear multidimensional equations (like the $\phi^4$-model) have.

In this article we shall consider the $(1+D)$ $|\phi|^4$-model in Minkowski space (summation
is implied over the repeating contravariant and covariant indices):
\begin{equation}
\partial_{\mu}\partial^{\mu}\phi+m^2\phi+\lambda|\phi|^2\phi=0,
\label{equation}
\end{equation}
with metric
$$
g_{\mu\nu}={\rm diag}(+1,-1,-1,...,-1).
$$
As we shall see, in the case $D=2$ for the negative coupling
$\lambda$ the equation (\ref{equation}) admits the nonsingular
solutions $\phi(x^{\mu})$, $\mu=0,\,1,\,2$ such that
$$
|\phi(x^{\mu})|\to B={\rm const}\qquad {\rm at}\qquad
x^2+y^2\to\infty.
$$
Of course, at first sight models with the negative coupling are quite suspicious from the physical point of view.
In fact, since the potential is negative, the action will not be bounded from below and
therefore the model may not be stable quantum-mechanically. On the other hand, we have learned recently that the models
with negative potentials might be extremely important in cosmology and strings theory
\cite{Linde}. Thus, the problem to find out the interesting exact solutions of such
models is alive and kicking.

We would also like to note, that it is still unclear, whether similar results can also be obtained for the case $D>2$, or not. We believe in former, but for now it is just a hypothesis.

\section{Equations}
In \cite{Yurov} there has been obtained the novel exact solution of the Davey-Stewartson II
(DS-II) equations describing the soliton on the plane wave background (see also \cite{BLP}). This solution was constructed via Darboux transformations  
\cite{DT}, \cite{DT-1} and has the form
\begin{equation}
\phi(x,y,t)=B{\rm
e}^{is(x,y,t)}\left(-1+\frac{P_1(x,y,t)}{P_2(x,y,t)}\right),
\label{anzatz}
\end{equation}
where $s(x,y,t)$ and $P_1(x,y,t)$ are linear functions whereas $P_2(x,y,t)$
is the polynomial of the second order such that $P_2(x,y,t)>0$ for all values of $x$, $y$, $t$.

The aim of this work is to show that the nonintegrable (1+2)
$|\phi|^4$ with negative coupling admits the similar  solution (see
(\ref{anzatz})) with
\begin{equation}
\begin{array}{l}
P_1=a_{\mu}x^{\mu}+a,\qquad P_2=\eta_{\mu\nu}x^{\mu}x^{\nu}+b_{\mu}x^{\mu}+A^2,\\
\\
s=s_{\mu}x^{\mu},
\end{array}
\label{forma}
\end{equation}
where
$$
\begin{array}{l}
\eta_{\mu\nu}=\eta_{\nu\mu},\qquad
\eta_{\mu\nu}=\left(\eta_{\mu\nu}\right)^*,\\
\\
(A^2)^*=A^2, \qquad B^*=B,\qquad (s_{\mu})^*=s_{\mu}.
\end{array}
$$
Substituting (\ref{forma}) into the (\ref{equation}) leads to
\begin{equation}
\begin{array}{l}
iP_2\left[P_2\left(P_1-P_2\right)\partial_{\mu}\partial^{\mu}s+2J^{\mu}\partial_{\mu}s\right]+
P_2\partial_{\mu}J^{\mu}-2J^{\mu}\partial_{\mu}P_2+\\
\\
+\left(P_1-P_2\right)\left[\lambda
B^2(P_1-P_2)(P_2-P^*_1)+(m^2-\partial_{\mu}s\partial^{\mu}s)P_2^2\right]=0,
\end{array}
\label{uh}
\end{equation}
where
$$
J^{\mu}=P_2\partial^{\mu}P_1-P_1\partial^{\mu}P_2.
$$
Using (\ref{uh}) in the particular case $b_{\mu}=0$ will
result in the following system:
\begin{equation}
\begin{array}{l}
s_{\mu}s^{\mu}-\lambda B^2-m^2=0,\\
\\
(2is_{\mu}a^{\mu}+\lambda B^2(a+a^*))A^4-a(\lambda
B^2(a+2a^*)+2\eta^{\mu}_{\mu})A^2+\lambda B^2|a|^2a=0,\\
\\
a_{\mu}+a^*_{\mu}=0,\\
\\
\left(2is_{\mu}a^{\mu}+\lambda
B^2(a+a^*)\right)\eta_{\alpha\beta}-4is^{\mu}\eta_{\mu\alpha}a_{\beta}+\lambda
B^2a_{\alpha}a_{\beta}=0,\\
\\
2\eta_{\alpha\beta}\left[\left(\lambda B^2
a^*+\eta^{\mu}_{\mu}\right)a_{\rho}+2\left(a_{\mu}+ias_{\mu}\right)\eta^{\mu}_{\rho}\right]+
a_{\rho}\left[\lambda B^2
a_{\alpha}a_{\beta}-8\eta_{\mu\alpha}\eta^{\mu}_{\beta}\right]=0,\\
\\
\left[\left(a(+2a^*)-2A^2(a+a^*)\right)\lambda
B^2+8iaA^2\eta^{\mu}_{\mu}s^{\nu}a_{\nu}\right]\eta_{\alpha\beta}-\lambda
B^2\left(A^2+a^*-2a\right)a_{\alpha}a_{\beta}-\\
-8a\eta_{\mu\alpha}\eta^{\mu}_{\beta}+4iA^2s^{\mu}\eta_{\mu\alpha}a_{\beta}=0,\\
\\
\left[\lambda
B^2\left(2a^*A^2-2|a|^2+a^2\right)+2\eta^{\mu}_{\mu}A^2\right]a_{\alpha}+4A^2\left(a^{\mu}+ias^{\mu}\right)\eta_{\mu\alpha}=0.
\end{array}
\label{sys}
\end{equation}
Thus in the case $(1+D)$ one has to solve the system of
$$
N(D)=\frac{(D+2)(2D^2+5D+7)}{2}
$$
algebraic equations, provided that:
\newline
(i) $P_2>0$ for any $x^{\mu}$ (if this is the case then solution
(\ref{anzatz}) will be nonsingular) and
\newline
(ii) the level lines of the function $P_2$ will be closed curves.

\section{$D=2$ Solutions}

In the case $D=2$: $N(D)=50$. The  inequalities (i) shall be valid
if
\begin{equation}
\begin{array}{c}
\eta_{11}>0,\qquad
\left|\begin{array}{cc} \eta_{11}&\eta_{12}\\
\eta_{12}&\eta_{22}
\end{array}\right|>0,\\
\\
\left|\begin{array}{ccc} \eta_{00}&\eta_{01}&\eta_{02}\\
\eta_{01}&\eta_{11}&\eta_{12}\\
\eta_{02}&\eta_{12}&\eta_{22}
\end{array}\right|>0,\qquad
\left|\begin{array}{cccc} 0&b_{0}&b_{1}&b_{2}\\
b_{0}&\eta_{00}&\eta_{01}&\eta_{02}\\
b_{1}&\eta_{01}&\eta_{11}&\eta_{12}\\
b_{2}&\eta_{02}&\eta_{12}&\eta_{22}
\end{array}\right|>0.
\end{array}
\label{nerav}
\end{equation}
Let $\eta_0$, $\alpha$, $\beta$, $\gamma$, $\rho$ and $b$ be the
arbitrary real parameters. Lets define three Lorentzian vectors
$$
\xi_{\mu}=(\xi_0,\xi_1,\xi_2),\qquad
\eta_{\mu}=(\eta_0,\eta_1,\eta_2),\qquad
\theta_{\mu}=(\theta_0,\theta_1,\theta_2),
$$
such that
\begin{equation}
\begin{array}{l}
\displaystyle{ \eta_1=\eta_0\cos\,\alpha,\qquad
\eta_2=\eta_0\sin\,\alpha,\qquad
\sigma=\frac{\rho^2\sin[2(\beta-\alpha)]}{4\eta_0\sin(\alpha-\gamma)},}\\
\\
\xi_1=\rho\cos\,\beta,\qquad \xi_2=\rho\sin\,\beta,\qquad
\xi_0=\rho\cos(\beta-\alpha),\\
\\
\displaystyle{ \theta_1=\sigma\cos\,\gamma,\qquad
\theta_2=\sigma\sin\,\gamma,\qquad
\theta_0=\frac{\sigma\cos(\beta-\gamma)}{\cos(\beta-\alpha)}, }
\end{array}
\label{vectors}
\end{equation}
and, finally
\begin{equation}
\lambda=-2\rho^2\sin^2(\alpha-\beta),\qquad
m^2=\frac{\sigma^2\sin(2\beta-\gamma-\alpha)\sin(\gamma-\alpha)}{\cos^2(\beta-\alpha)}.
\label{couplings}
\end{equation}
Then the solution of the system (\ref{sys}), such that
(\ref{nerav}) together with the conditions (i) and (ii) holds has the
form:
\begin{equation}
\begin{array}{l}
\eta_{\mu\nu}=\xi_{\mu}\xi_{\nu}-2b\xi_{\mu}\eta_{\nu}+4\left(b^2+B^2\right)
\eta_{\mu}\eta_{\nu},\qquad
s_{\mu}=\left(2B^2-b^2\right)\eta_{\mu}+b\xi_{\mu}+\theta_{\mu},\\
\\
\displaystyle{ a_{\mu}=4i\eta_{\mu},\qquad a=\frac{1}{B^2},\qquad
A=\pm\frac{1}{2B}.}
\end{array}
\label{solution1}
\end{equation}
This is true when $b_{\mu}=0$. If $b_{\mu}\ne 0$ then, instead of (\ref{sys}) one will have
a somehow more difficult system. The particular solution, however, can still be obtained:
\begin{equation}
\begin{array}{l}
\displaystyle{
\eta_{\mu\nu}=4B^2\left(\xi_{\mu}\xi_{\nu}-2b(\xi_{\mu}\eta_{\nu}+\xi_{\nu}\eta_{\mu})+
4\left(b^2+B^2\right)\eta_{\mu}\eta_{\nu}\right),}\\
\\
\displaystyle{b_{\mu}=8B^2\left(2(\kappa\psi B+\chi
b)\eta_{\mu}-\chi\xi_{\mu}\right),\qquad
a_{\mu}=16iB^2\eta_{\mu},}
\\
\\
\displaystyle{ a=4\left(1+\frac{\kappa
B}{|c_1|^2}(c_2c^*_1-c^*_2c_1)\right),\qquad A^2=4B^2(\chi^2+\psi^2),}\\
\\
\displaystyle{
\chi=\frac{\kappa|c_1|^2-B(c_1c^*_2+c^*_1c_2)}{2B|c_1|^2},\qquad
\psi=\frac{i(c_1c^*_2-c^*_1c_2)}{2|c_1|^2},}
\end{array}
\label{solution2}
\end{equation}
where $c_{1,2}$, $\kappa=\pm 1$ are arbitrary complex constants and
$s_{\mu}$ is the same as in the (\ref{solution1}).

Finally, it is easy to show that
$$
\eta_{\mu\nu}x^{\mu}x^{\nu}+b_{\mu}x^{\mu}+A^2=1+4B^2\left[\left(\Lambda_{\mu}x^{\mu}-
\chi\right)^2+\left(\L_{\mu}x^{\mu}+ \psi\right)^2\right],
$$
where
$$
\Lambda_{\mu}=\xi_{\mu}-2b\eta_{\mu},\qquad L_{\mu}=2\kappa
B\eta_{\mu},
$$
which means that (\ref{nerav}) and two conditions (i) and (ii)
hold for our solution, and that concludes our proof.
\section{Conclusion}

As we have seen, the equation (\ref{equation}) with negative coupling
$\lambda$ admits the nonsingular solution which looks as the linear
superposition of the plane wave and rational soliton, and the level lines
of this solution on the plane $xy$ are ellipses.

There remains two open questions, both of them being an interesting problem for the future
investigations.

The first question is the possible generalization of the method to the case
$D>2$. As for now, we can't see any possible reasons why our approach will not work
in multidimensions. The only price we have to pay working there is a rapidly
growing amount of calculations. In fact, the number of
algebraic equations $N(D)$, having to be solved grows as $D^3$:
$N(3)=100$, $N(4)=177$ and $N(10)=1542$. But then again: the usage of computer programs (for example, MAPLE), that are powerful enough to handle such systems in a reasonable periods of time allows for this problem to be solved.

The second question is the negative value of $\lambda$. Is it possible to
construct the exact solution with $\lambda>0$? The answer is yes, but
in this case the solutions will be singular. The singularity will be located along the hyperbola and the physical
meaning of such solution is unclear for us.

At last, the next step is the calculation of one-loop quantum corrections to these new classical nontrivial solutions \cite{Bordag} but this is subject of another investigation. 

\section*{Acknowledgements}

We'd like to thank our son and colleague Valerian  Yurov for
useful comments and for the interest in this work.


 \end{document}